\documentclass[english,12pt]{article}
\usepackage[T1]{fontenc}
\usepackage[latin9]{inputenc}
\usepackage{amsthm}
\usepackage{amsmath}
\usepackage{graphicx}

\theoremstyle{plain}

\usepackage{babel}

\begin{document}

\title{Axigluon Couplings in the Presence of Extra Color-Octet Spin-One Fields}

\author{Alfonso R. Zerwekh\\
 Centro de Estudios Subatómicos and Instituto de Física\\
 Facultad de Ciencias\\
 Universidad Austral de Chile\\
 Casilla 567, Valdivia, Chile}

\date{}
\maketitle 
\begin{abstract}
In this paper, we study how the interaction of the axigluon with quarks
is modified when we introduce new color-octet spin-one fields in a
chiral color model. We show that in this case the strength of this
interaction is not completely determined by the gauge symmetry any
more and can be significantly weaker than the one predicted in the
original chiral color model. In this way, we reinterpret the non-observability
of the axigluon at the Tevatron, not as a limit on the axigluon mass,
but as a limit on the strength of the axigluon coupling to quarks. 
\end{abstract}

\section{Introduction}

It is inherent to science to look forward to new horizons and explore
every possible path in the hope to increase our knowledge about Nature.
That's why, despite the amazing phenomenological success of the Standard
Model, we are always tempted to extend its particle content or its
symmetry and study the properties of the resulting model. Of course
there are good theoretical reasons for doing it: we need to explain
many mysteries present in the foundations of the Standard
Model. Perhaps the best known example is the problem of naturalness
in the Higgs sector. The necessity of understanding why we observe a
so large hierarchy between the electroweak scale and, for example,
the Grand Unification scale, have motivated the construction of models
as different as supersymmetry, Technicolor or Little Higgs, among
others. In all these theories, the hierarchy problem is solved at
the price of significantly extend the particle and symmetry contents
proposed by the Standard Model and verified to the level of 1\% at
LEP.

Another intriguing, but much less popular, question is why we observe
a so different structure for the weak and strong interactions. The
weak interaction is profoundly chiral while the strong interaction
makes no difference between left and right fermions. In this context,
and with the aim of unification in their eyes, Frampton and Glashow 
\cite{Frampton:1987ut,Frampton:1987dn} proposed many years ago a
family of models where the local gauge symmetry of the strong interaction is
supposed to be $SU(3)_{L}\times SU(3)_{R}$ at high energies. Of course
at some large energy scale this symmetry 
must be broken to its diagonal group which is identified with the
usual QCD group: $SU(3)_{c}$. A general prediction of this kind of
models is the existence of a massive color-octet spin-one particle
usually called axigluon because it has an axial coupling to quarks.
The strength of this coupling is, as we will see, completely dictated
by the gauge symmetry and, in the simplest, model is equal to the
gluon-quark interaction. Of course the presence of such a particle
would introduce a large contribution in the dijet production cross
section at hadron colliders. Unfortunately, such a signal has not been
observed and stringent 
limits have been placed on the axigluon mass \cite{Aaltonen:2008dn}.
Phenomenological constrains have also been obtained from top-antitop
events \cite{Ferrario:2009bz}.

In this paper, we study how the interaction of the axigluon with quarks
is modified when we introduce new color-octet spin-one fields.
We will show that in this case the strength of
this interaction is not completely determined by the gauge symmetry
any more and can be significantly weaker than the one predicted in
the original chiral color model. In this way we can reinterpret the
non-observability of the axigluon at the Tevatron not as a limit on
the axigluon mass, but as a limit on the strength of the axigluon
coupling to quarks.

\section{Chiral Color and the Axigluon}

We start by recalling the origin of the axigluon in chiral color models.
Of course, this revision will be schematic and we will not discuss
important points such as the choise of anomaly free representations
for fermions\footnote{A detailed discussion of this topics can be
  found in \cite{Frampton:1987ut}, \cite{Frampton:1987dn} and
  \cite{FramptonBook}}. We will concentrate only in those aspects we
judge important for the axigluon phenomenology.

Let us, then, consider the following Lagrangian:

\begin{equation}
\mathcal{L}=-\frac{1}{2}\mathrm{tr}\left\{ G_{L\mu\nu}G_{L}^{\mu\nu}\right\} -\frac{1}{2}\mathrm{tr}\left\{ G_{R\mu\nu}G_{R}^{\mu\nu}\right\} +\frac{f^{2}}{2}\mathrm{tr}\left\{ D_{\mu}U^{\dagger}D^{\mu}U\right\} \label{eq:LagGauge}\end{equation}

where

\begin{eqnarray*}
G_{L\mu\nu} & = & \partial_{\mu}l_{\nu}-\partial_{\nu}l_{\mu}-ig_{L}\left[l_{\mu},\: l_{\nu}\right]\\
G_{R\mu\nu} & = & \partial_{\mu}r_{\nu}-\partial_{\nu}r_{\mu}-ig_{R}\left[r_{\mu},\: r_{\nu}\right]\\
D_{\mu}U & = & \partial_{\mu}U-ig_{L}l_{\mu}U+ig_{R}Ur_{\mu}\\
D_{\mu}U^{\dagger} & =
& \partial_{\mu}U^{\dagger}-ig_{R}r_{\mu}U^{\dagger}+ig_{L}U^{\dagger}l_{\mu}
\end{eqnarray*}
and $l_{\mu}$ ($r_{\mu}$) is the gauge field of $SU(3)_{L}$ ($SU(3)_{R}$).
Of course, the first two terms are invariant under $SU(3)_{L}\times SU(3)_{R}$
transformations and the third term is a nonlinear sigma model term we include
to describe the breaking down of the original local gauge symmetry to
$SU(3)_{c}$. Although it is straightforward to develop the model
considering different coupling constants for $SU(3)_{L}$ and $SU(3)_{R}$
\cite{Martynov:2009en} we
are going to limit ourselves to the case where $g_{L}=g_{R}=g$. We
do it in order to simplify our analysis and because this was the choice
made by authors of the original chiral color model motivated by unification
arguments\cite{Frampton:1987ut}.

In the unitary gauge ($U=1$), the third term of equation (\ref{eq:LagGauge})
gives rise to a nondiagonal mass matrix for the gauge bosons.When
we diagonalize the mass matrix, we found the following eigenvalues
and eigenvectors:

\begin{eqnarray*}
m_{G} & = & 0\\
m_{A} & = & \frac{gf}{\sqrt{2}}\end{eqnarray*}

\begin{eqnarray}
G_{\mu} & = & \frac{1}{\sqrt{2}}\left(l_{\mu}+r_{\mu}\right)\label{eq:G1}\\
A_{\mu} & = & \frac{1}{\sqrt{2}}\left(l_{\mu}-r_{\mu}\right)\label{eq:A1}\end{eqnarray}
 where $G_{\mu}$ is the gluon and $A_{\mu}$ is the axigluon.

The interaction Lagrangian for the quarks can be written as:

\begin{equation}
\mathcal{L}=\frac{1}{2}g\bar{\psi}l_{\mu}\gamma^{\mu}(1-\gamma_{5})\psi+\frac{1}{2}g\bar{\psi}r_{\mu}\gamma^{\mu}(1+\gamma_{5})\psi\label{eq:2}\end{equation}

Inverting equations (\ref{eq:G1}) and (\ref{eq:A1}) we can write
the Lagrangian in term of the physical fields and we obtain:

\begin{equation}
\mathcal{L}=\frac{g}{\sqrt{2}}\bar{\psi}G_{\mu}\gamma^{\mu}\psi+\frac{g}{\sqrt{2}}\bar{\psi}A_{\mu}\gamma^{\mu}\gamma_{5}\psi.\label{eq:3}\end{equation}

In this minimal model, the gluon-quark and axigluon-quark interaction
terms have the same coupling constant $g_{QCD}\equiv g/\sqrt{2}$.
In more general models, where $g_{L}\neq g_{R}$, it is not true and
 $g$ appears modulated by trigonometric functions of a mixing
angle\cite{Martynov:2009en}.

\section{Our Model}

Following the ideas previously developed by the author in \cite{Zerwekh:2003zz}
 and \cite{Zerwekh:2006te}, we
propose to modify the model described above by adding new spin-one
fields, $L_{\mu}$ and $R_{\mu}$, which transform like gauge fields
under $SU(3)_{L}$ and $SU(3)_{R}$ respectively with a characteristic
coupling constant $g'$. The Lagrangian that describes the gauge sector
of the model, including the effective symmetry breaking term, is

\begin{eqnarray}
\mathcal{L} & = & -\frac{1}{2}\mathrm{tr}\left\{ G_{L\mu\nu}G_{L}^{\mu\nu}\right\} -\frac{1}{2}\mathrm{tr}\left\{ G_{R\mu\nu}G_{R}^{\mu\nu}\right\} \notag\\
 &  & -\frac{1}{2}\mathrm{tr}\left\{ \rho_{L\mu\nu}\rho_{L}^{\mu\nu}\right\} -\frac{1}{2}\mathrm{tr}\left\{ \rho_{R\mu\nu}\rho_{R}^{\mu\nu}\right\} \notag\\
 &  & +\frac{M^{2}}{g'^{2}}\mathrm{tr}\left\{ \left(gl_{\mu}-g'L_{\mu}\right)^{2}\right\} +\frac{M^{2}}{g'^{2}}\mathrm{tr}\left\{ \left(gr_{\mu}-g'R_{\mu}\right)^{2}\right\} \notag\\
 &  & +\frac{f^{2}}{2}\mathrm{tr}\left\{
   D_{\mu}U^{\dagger}D^{\mu}U\right\} \label{eq:LagMyModel}
\end{eqnarray}
where $G_{L\mu\nu}$, $G_{R\mu\nu}$ and $U$ are the same than in
the previous section while $\rho_{L\mu\nu}$ and $\rho_{R\mu\nu}$
are defined by

\begin{eqnarray*}
\rho_{L\mu\nu} & = & \partial_{\mu}L_{\nu}-\partial_{\nu}L_{\mu}-ig'\left[L_{\mu},\: L_{\nu}\right]\\
\rho_{R\mu\nu} & = & \partial_{\mu}R_{\nu}-\partial_{\nu}R_{\mu}-ig'\left[R_{\mu},\: R_{\nu}\right]\end{eqnarray*}

Again, in the unitary gauge a nondiagonal mass matrix is explicitly
generated. Although the resulting mass matrix can be exactly diagonalized,
in order to obtain easily manipulable expressions we will assume that
$g'\gg g$ and we will write our results in the first order in $g/g'$.
The eigenvalues, that is, the masses of the physical states are:

\begin{eqnarray*}
m_{G} & = & 0\\
m_{A} & = & \frac{gf}{\sqrt{2}}\\
m_{G'} & = & M\\
m_{A'} & = & M\end{eqnarray*}

Thus, the physical spectrum is composed of a (exactly) massless gluon,
an axigluon with the same mass (in this limit) than in the minimal
model and two degenerate (at this level of approximation) heavy gluon
and axigluon. Notice that the new mass scale $M$ is not constrained
and we can safely suppose that it is large enough to prevent the observation of the heavy states.

The normalized mass eigenvectors can be written as:

\begin{eqnarray}
G_{\mu} & = & \frac{1}{\sqrt{2}}l_{\mu}+\frac{1}{\sqrt{2}}r_{\mu}+\frac{g}{\sqrt{2}g'}L_{\mu}+\frac{g}{\sqrt{2}g'}R_{\mu}\notag\\
A_{\mu} & = & -\frac{1}{\sqrt{2}}l_{\mu}+\frac{1}{\sqrt{2}}r_{\mu}-\frac{g}{\sqrt{2}g'}\left(1-\frac{m_{A}^{2}}{M^{2}}\right)L_{\mu}+\frac{g}{\sqrt{2}g'}\left(1-\frac{m_{A}^{2}}{M^{2}}\right)R_{\mu}\notag\\
G'_{\mu} & = & \frac{g}{\sqrt{2}g'}l_{\mu}+\frac{g}{\sqrt{2}g'}r_{\mu}-\frac{1}{\sqrt{2}}L_{\mu}-\frac{1}{\sqrt{2}}R_{\mu}\notag\\
A'_{\mu} & = &
-\frac{g}{\sqrt{2}g'}\left(1-\frac{m_{A}^{2}}{M^{2}}\right)^{-1}l_{\mu}+\frac{g}{\sqrt{2}g'}\left(1-\frac{m_{A}^{2}}{M^{2}}\right)^{-1}r_{\mu}+\notag\\
& &+\frac{1}{\sqrt{2}}L_{\mu}-\frac{1}{\sqrt{2}}R_{\mu} \label{eq:estadosfisicos}
\end{eqnarray}

Because we have now two fields that transform as gauge fields for
each group ($l_{\mu}$ and $L_{\mu}$ for $SU(3)_{L}$ and $r_{\mu}$
and $R_{\mu}$ for $SU(3)_{R}$) any combination of the form
$g(1-k)l_{\mu}+g'kL_{\mu}$ 
and $g(1-k')r_{\mu}+g'k'R_{\mu}$, where $k$ and $k'$ are arbitrary
constants, can be used to construct covariant
derivatives\cite{Zerwekh:2003zz}. This means 
that the Lagrangian describing the gauge interaction of quarks can
be written as:

\begin{eqnarray}
\mathcal{L} & = & \frac{1}{2}g(1-k)\bar{\psi}l_{\mu}\gamma^{\mu}(1-\gamma_{5})\psi+\frac{1}{2}g'k\bar{\psi}L_{\mu}\gamma^{\mu}(1-\gamma_{5})\psi+\nonumber \\
 &  & +\frac{1}{2}g(1-k')\bar{\psi}r_{\mu}\gamma^{\mu}(1+\gamma_{5})\psi+\frac{1}{2}g'k'\bar{\psi}R_{\mu}\gamma^{\mu}(1+\gamma_{5})\psi\label{eq:MyLagFerm}\end{eqnarray}
In principle, $k$ and $k'$ are independent parameters but, again,
for simplicity we will assume that $k=k'$. Using Lagrangian (\ref{eq:MyLagFerm})
and the definition of the physical fields, we can obtain the terms
of the Lagrangian that couple the gluon and the axigluon to quarks 

\begin{equation}
\mathcal{L}=\frac{g}{\sqrt{2}}\bar{\psi}G_{\mu}\gamma^{\mu}\psi+
\frac{g}{\sqrt{2}}\left(1-\chi 
\right)\bar{\psi}A_{\mu}\gamma^{\mu}\gamma_{5}\psi
\end{equation} 
where $\chi$ is defined as:

\begin{equation}
  \label{eq:chi}
  \chi \equiv \frac{m^2_A}{M^2}k
\end{equation}

Notice that, while the coupling of the gluon to the quarks is the
usual one and is independent of $k$ (because is protected by the
$SU(3)_{c}$ gauge symmetry), the coupling constant of the axigluon
has been modified and now depend on the free parameter $k$. Of course
we can use this freedom to choice an adequated valued $k$ in order to make the axigluon invisible at the Tevatron.

Let us briefly comment about to keep $k$ and $k'$ as independent
parameters. In this case, additionally to the axial coupling, a new
vector coupling would appear between the axigluon and the quarks,
proportional to $(k'-k)$.

\section{Results}

Now we want to estimate the values of $\chi$ 
allowed by experiments. It is evident that, in order to keep the axigluon
invisible in dijet observations at the Tevatron, the following inequality
must be satisfied:

\begin{equation}
\left(1-\chi \right)^{2}\sigma_{A}\leq\sigma_{95\%}\label{eq:limites}
\end{equation}
where $\sigma_{95\%}$ is the maximum value of the production cross
section of a resonance decaying into dijets, allowed by experiment
at 95\% C.L. and $\sigma_{A}$ is the cross section for the production
of an axigluon and its decay into dijets in the usual minimal model
($k=0$).

We use Calchep \cite{Pukhov:2004ca}  in order to estimate $\sigma_{A}$
for $\sqrt{s}=1.96$ 
TeV, and $\left|y\right|\leq1$ (where $y$ is the rapidity of the
jet) for values of the axigluon mass in the interval $\left[260,\:1000\right]$
GeV. For $\sigma_{95\%}$, we use the values reported in \cite{Aaltonen:2008dn}.

\begin{figure}
\begin{centering}
\includegraphics[scale=0.8]{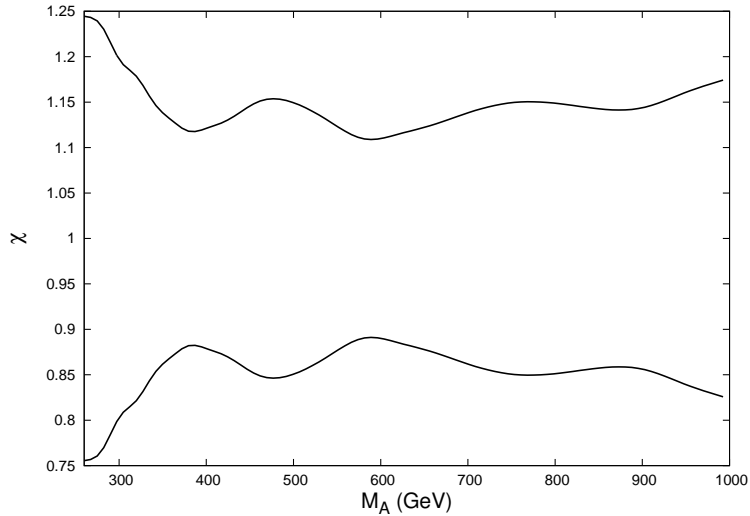}\label{Flo:Limites}
\par\end{centering}

\caption{Limits for free parameter $\chi$ defined in equation
 (\ref{eq:chi}). The region between the continuous lines is allowed. } 

\end{figure}

The results are summarized in figure \ref{Flo:Limites}. The continuous
curves represent the limits on $\chi$ obtained
using the equality in (\ref{eq:limites}). The allowed region is the
one between the curves. For $\chi=1$, the axigluon completely decouple from quarks but, we can see, a deviation of $\chi$ from $1$
of about $11\%$ is allowed for the whole mass range considered
here. But for low masses a deviation as large as $25\%$ is still possible.

\section{Conclusion}

In this work we have presented a new mechanism to reduce the coupling
of the axigluon to quarks in order to evade the limits imposed by
dijet production experiments. This mechanism consists in adding new
spin-one fields that originally transform like gauge fields under
$SU(3)_{L}$ and $SU(3)_{R}$ respectively and construct with them
generalized covariant derivatives. We want to emphasize that our
method is rather innocuous. Our model preserve the fundamental
structure of chiral color models and from equation
(\ref{eq:estadosfisicos}) we can see that the descomposition of the
physical states in terms of the original has the same structure for
the light fields (the gluon and the axigluon) and the heavy ones. That
implies that their interaction to quarks has also the same
structure. Thus, we expect that once one has chosen anomaly free
representations  for quarks in the minimal model, the potentially
new anomalous diagrams introduced in our model by the heavy states
will also cancel. On the other hand, although we have shown that our
proposal works in the important case where we
keep the symmetry between the left and right sectors, it is
straightforward to generalize it to the case when $g_L \neq g_R$. In
this case, we can reduce independently the axial  as well as the
vector coupling that the axigluon develops in this generalized
models. For doing that, 
it is enough to keep $k \neq k'$ in equation (\ref{eq:MyLagFerm}).
Finally, the appearance of a heavy gluon and a 
heavy axigluon is not problematic from a phenomenological point of
view since the model allows their mass to be large enough to be invisible at the
Tevatron.

\section*{Acknowledgement}

A. R. Z. is partially supported by Fondecyt grant 1070880. The author
wants to thank the hospitality of the Instituto de Física Teórica
(Unesp), S\~ao Paulo, Brazil, where this work was completed.TGD

\end{document}